\documentclass[twocolumn,preprintnumbers,amsmath,amssymb]{revtex4}

\usepackage{ifthen}
\usepackage{ifpdf}
\usepackage{color}

\ifpdf
\usepackage{graphicx}
\usepackage{epstopdf}
\else
\usepackage{graphicx}
\usepackage{epsfig}
\fi

\graphicspath{{./}{./Figs/}}

\usepackage{latexsym}
\usepackage{amsmath}
\usepackage{amssymb}
\usepackage{bm}
\usepackage{wasysym}

\newcommand{\trc}{\mbox{trace}}

\newcommand{\eexp}{\mbox{e}^}

\newcommand{\be}[1]{\begin{eqnarray}\ifthenelse{#1=-1}{\nonumber}{\ifthenelse{#1=0}{}{\label{e#1}}}}
\newcommand{\beq}{\begin{eqnarray}}
\newcommand{\eeq}{\end{eqnarray}} 

\newcommand{\mpg}[2][1.0\hsize]{\begin{minipage}[b]{#1}{#2}\end{minipage}}

\newcommand{\hide}[1]{\textcolor{red}{[hidden text]}}

\newcommand{\Eq}[1]{\textcolor{blue}{Eq.\!\!~(\ref{#1})}} 
\newcommand{\Fig}[1]{\textcolor{blue}{Fig.}\!\!~\ref{#1}}

\newcommand{\mycite}[1]{\textcolor{blue}{\cite{#1}}}

\begin{document}

\title{Stability and stabilization of unstable condensates}

\author{Doron Cohen}

\affiliation{Departments of Physics, Ben-Gurion University of the Negev, Beer-Sheva 84105, Israel}

\date[]{Physica Scripta {\bf T165}, 014032 (2015). Special issue. Proceedings of FQMT conference (Prague, 2013).}

\begin{abstract}
It is possible to condense a macroscopic number of bosons 
into a single mode. Adding interactions the question arises 
whether the condensate is stable. For repulsive interaction 
the answer is positive with regard to the ground-state, 
but what about a condensation in an excited mode? 
We discuss some results that have been obtained for  
a 2-mode bosonic Josephson junction, and for a 3-mode 
minimal-model of a superfluid circuit. Additionally we 
mention the possibility to stabilize an unstable condensate  
by introducing periodic or noisy driving into the system: 
this is due to the Kapitza and the Zeno effects.  
\end{abstract}



\maketitle

\section{Introduction}

\begin{figure}[b!]
\centering

\includegraphics[height=0.25\hsize]{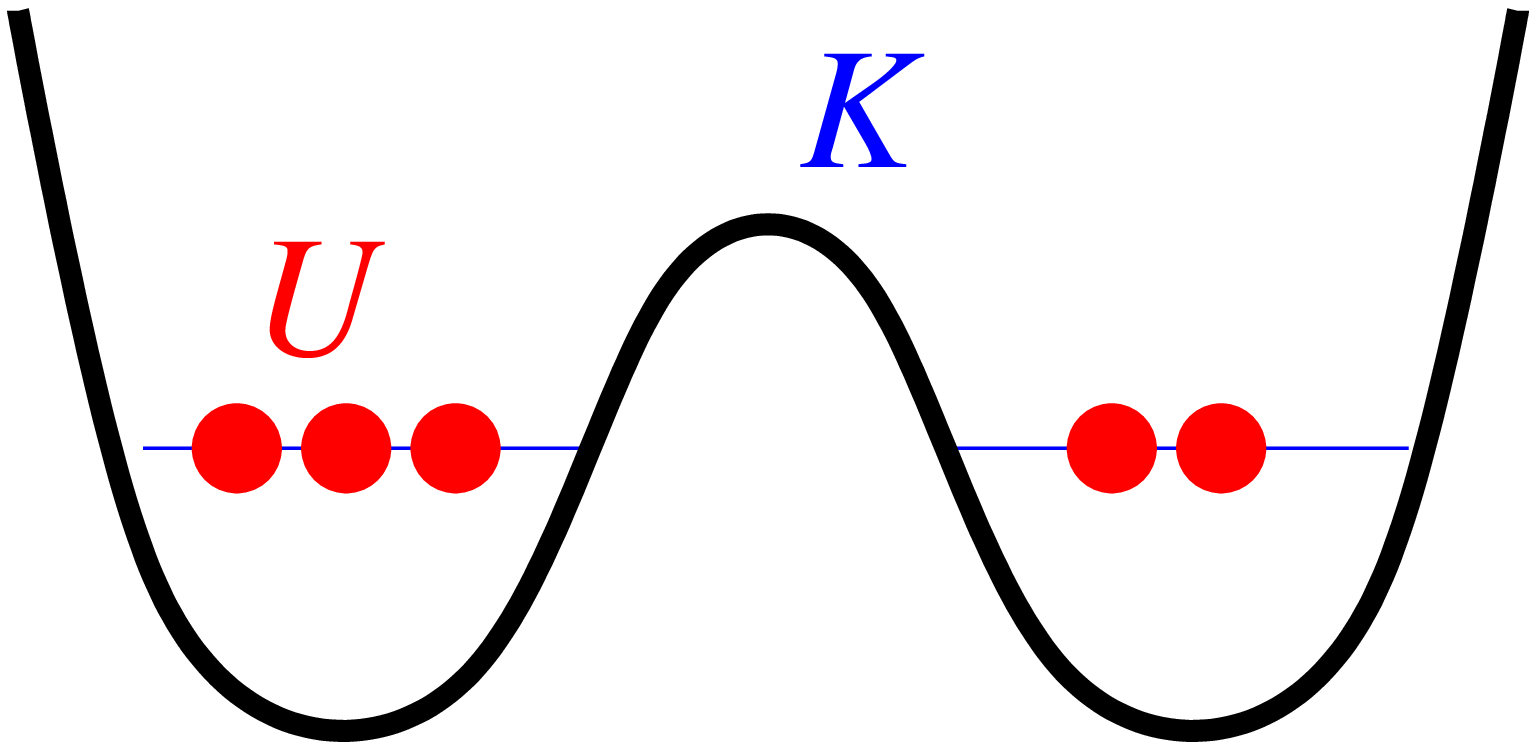}
\ \ \ \ \ \ \ \ 
\includegraphics[height=0.35\hsize]{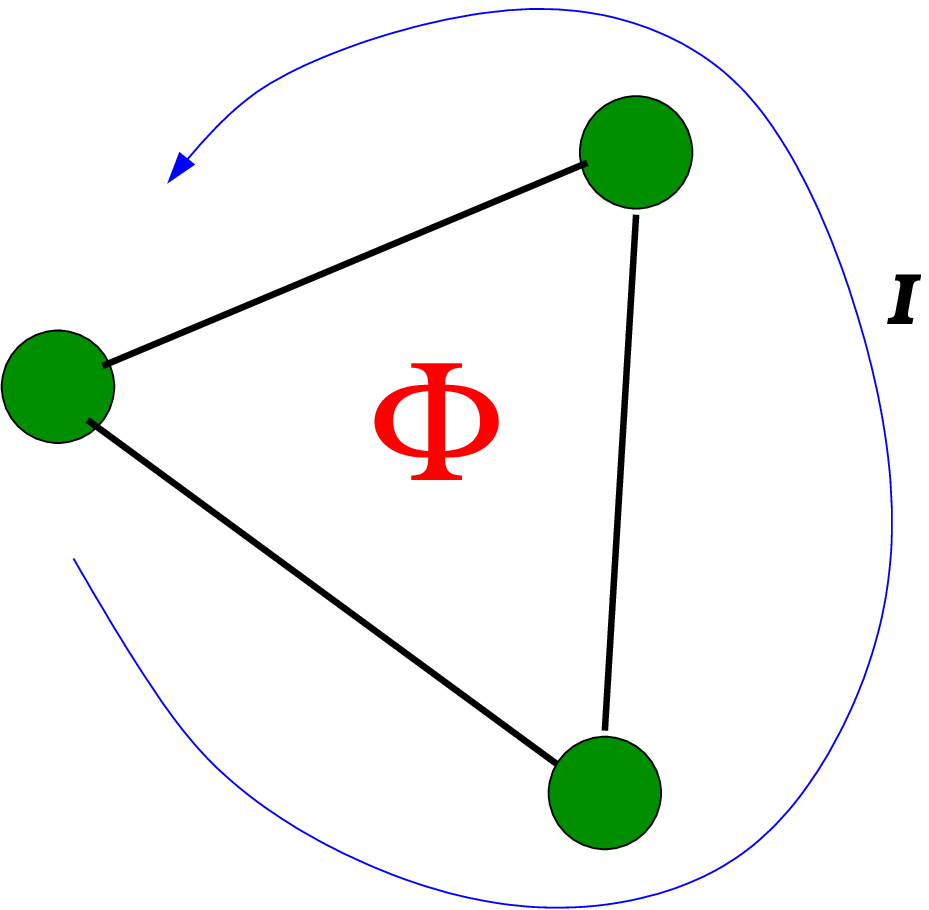}

\caption{
Illustrations of the dimer (left) and of the trimer (right) model systems.
Namely, we consider $N$ bosons that are described by an $M=2$ or by an $M=3$ site BHH.
The hopping frequency is $K$, and the on-site interaction is $U$.
In the case of a trimer the hopping frequencies acquire phases 
whose sum $\Phi$ reflects the Coriolis force.       
} 

\label{f1}
\end{figure}

This presentation concerns a system of $N$ spinless bosons 
in an $M$ site system, that are described 
by the Bose-Hubbard Hamiltonian (BHH) \mycite{BHH1}.
The explicit form the BHH will be provided in later sections.
At this stage of the introduction it is enough to say that 
the bosons can hop from site to site with hopping frequency~$K$, 
and that additionally there is an on-site interaction~$U$.
Accordingly the dimensionless interaction parameter is  
\beq
u \ \ = \ \ \frac{NU}{K}
\eeq
The model systems of interest are illustrated in \Fig{f1}.  
We refer to the $M=2$ system as the ``dimer" or as a bosonics Josephson junction. 
We refer to the $M=3$ system as the ``trimer" \mycite{trimer5,trimer11}  
or as a minimal model for a superfluid circuit \mycite{Udea,Altman,Carr1,Brand2,Ghosh}. 
In the latter case there appears in the BHH 
an additional dimensionless parameter $\Phi$ 
that reflects the rotation frequency of the device.    

The term ``orbital" is used in order to refer to 
a single particle state. The momentum orbitals 
of the $M$-site model systems of \Fig{f1} are   
\beq
|\varphi \rangle \ \ =  \ \ \frac{1}{\sqrt{M}}\sum_{j=1}^M \ [\eexp{i \varphi}]^j \ |j\rangle 
\eeq
These are the eigenstates of a single particle in the system.
The dimer has a lower mirror-symmetric orbital ${\varphi=0}$, 
and an upper anti-symmetric orbital ${\varphi=\pi}$. 
The momentum eigenstates of the trimer are ${\varphi=(2\pi/3)m}$, 
with ${m=0,\pm1}$.

Strict condensation means to place all the bosons in a single orbital.
Condensation in a momentum-orbital of the trimer is  
known as vortex-state. Condensation in a single site-orbital 
is known as self-trapped or as bright-soliton state.  
More generally we shall characterize the eigenstates 
of the BHH by a purity measure ${S\in[0,1]}$. 
Namely, given an eigenstate we define the reduced one-body 
probability matrix  $\rho_{ij}=(1/N)\langle a_j^{\dag} a_i\rangle$, 
where $a_j^{\dag}$ are the creation operators.  
From that we calculate $S \equiv \trc(\rho^2)$. 
Accordingly $S=1$ implies condensation in a single 
orbital, also termed ``coherent state", 
while ${1/S \sim M}$ implies a maximally fragmented state. 
The purity measure $S$ reflects the one-body coherence 
of the many-body state: small value of $S$ implies loss 
of fringe visibility in an interference experiment.

The condensation of all the bosons in a single $\varphi$ orbital 
is an eigenstate of the BHH in the absence of interaction. 
The other many-body eigenstates are fragmented,  
meaning that several orbitals are populated. 
Once we turn-on the interaction, the possible scenarios
are as follows: 
{\bf (1)}~The interaction {\em stabilizes} the $\varphi$  condensate. 
{\bf (2)}~As $u$ is varied a {\em bifurcation} is induced, 
such that the $\varphi$ condensate becomes unstable, 
and instead we get $M$ stable self-trapped states. 
{\bf (3)}~The interaction {\em mixes} the unperturbed $\varphi$ state 
with other fragmented unperturbed eigenstates. 
In the latter case we shall distinguish between: 
{\bf (3a)}~a quantum ``Mott transition" scenario; 
and   
{\bf (3b)}~a semiclassical ``ergodization" scenario.   
The last possibility is relevant if the underlying 
phase-space is chaotic.

In the next sections we shall discuss the stability 
of the $\varphi$ condensates. The outline is as follows: 
In section~II we discuss the simplest examples for 
quantum quasi-stability and quantum scarring \mycite{scars2}. 
In section~III we explain that an unstable state 
can be semiclassically stabilized by introducing 
high frequency periodic driving or noise into the system.
In section~IV we consider the trimer system, 
and discuss the possibility to witness a metastable 
vortex-state.
In the latter context we would like to clarify 
that the essence of ``superfluidity" is the possibility 
to witness a metastable vortex-state.
The term ``metastability"  rather than ``stability" 
indicates that the eigenstate is located in an intermediate energy range. 
Using a simple-minded phrasing this implies that there is 
a possibility to observe a current-carrying 
stationary-state that is not decaying. The stability 
of such stationary state is due to the interaction. 
This presentation is based on \mycite{csd,csf,ckt,csk,csm,sfs,sfc}
and further references therein \mycite{more}.

\section{The dimer - a minimal model for self trapping and Mott transition}

The BHH of an $M$ site system is
\be{3}
\mathcal{H} \ = \  
\frac{U}{2} \sum_{j=1}^{M}  a_{j}^{\dag} a_{j}^{\dag} a_{j} a_{j} 
 - \frac{K}{2} \sum_{j=1}^{M} \left(a_{j{+}1}^{\dag} a_{j} + a_{j}^{\dag} a_{j{+}1} \right)
\ \ \ \ 
\eeq
where $j=1\cdots M$ is the site index, $a_j^{\dag}$ are the 
creation operators, and $n_j=a_j^{\dag}a_j$ are the occupation operators. 
The total number of particles ${N=n_1+n_2}$ is a constant 
of motion hence the dynamics of an $M=2$ dimer is reduced 
to that of one degree-of-freedom 
\be{4}
J_z \ \ = \ \ n \ \ = \ \ \frac{1}{2}(n_1-n_2)
\eeq
An optional way to write the dimer Hamiltonian is 
to say that $J_z$ is like the $Z$ component of a $j=N/2$ 
spin entity. Using this language the hopping term of the BHH 
merely generates Rabi rotations around the X axis.
This means that the population oscillates between the two wells. 
The full BHH, including the interaction term, is written as follows
\be{5}
\mathcal{H}_{\text{dimer}} \ = \ U \hat J_z^2  \ - \  K \hat J_x  
\eeq
Semiclassically the spin orientation is described 
by the conjugate coordinates ${(\theta,\varphi)}$, 
or equivalently by ${(n,\varphi)}$, where $n=(N/2)\cos(\theta)$.   
With each point in phase-space we can associate 
a spin-coherent-state $|n,\varphi\rangle$ 
that is obtained by SU(2) rotation of the North-pole condensation 
state ${(a_1^{\dag})^N|\text{vaccum}\rangle}$.  
An arbitrary quantum state can be represented 
by the Husimi phase-space distribution 
${\rho(\varphi,n) = |\langle n,\varphi | \psi(t) \rangle|^2 }$.
In the following paragraphs we describe how the $UJ_z^2$ term 
of the dimer Hamiltonian affects the Rabi rotations that are 
generated by $KJ_x$.  

\begin{figure}
\centering

\includegraphics[width=0.8\hsize]{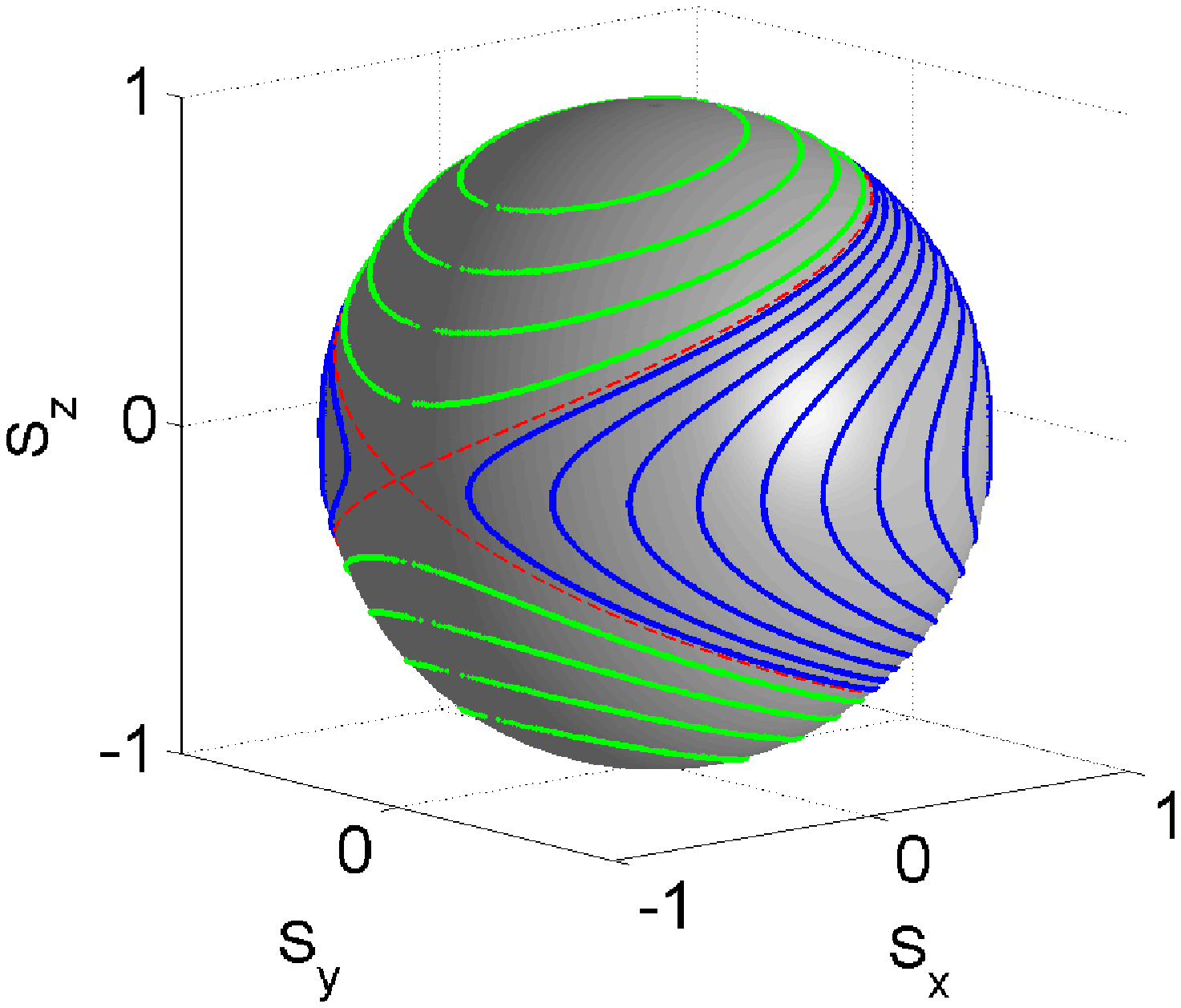}

\includegraphics[width=0.8\hsize]{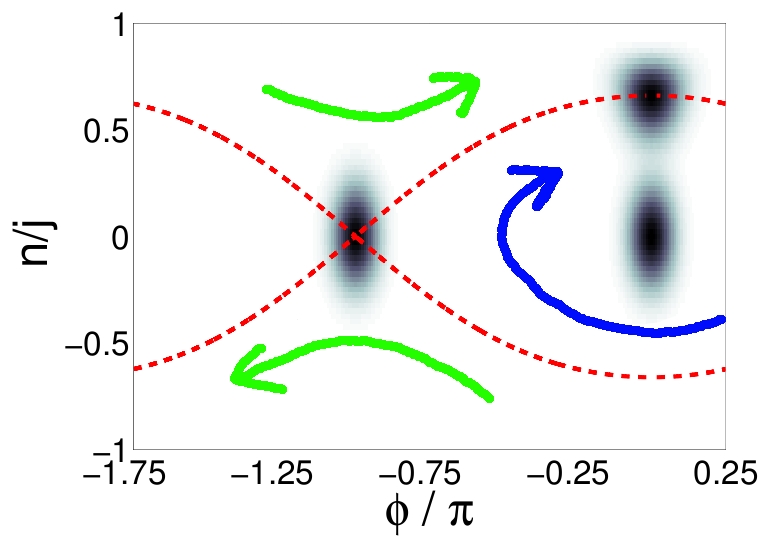}

\caption{
The dimer BHH is formally like that of a spin $j=N/2$ entity. 
Its spherical phase-space ${(\theta,\varphi)}$ is illustrated  
in the upper panel in the case ${u>1}$. It is similar 
to the cylindrical phase-space ${(n,\varphi)}$ of a mathematical 
pendulum (lower panel).  
We have 3 types of motion separated by a separatrix: 
Rabi oscillations between the two wells (blue curves); 
and self-trapped motion either in the left 
or in the right well (green curves). 
In the lower panel the 3 shaded Husimi distributions
represent the following preparations: 
low-energy $\varphi{=}0$ coherent-state;  
separatrix-energy $\varphi{=}\pi$ coherent-state;
and another coherent-state with the same energy.  
} 

\label{f2}
\end{figure}

{\bf Bifurcation scenario.--} 
For ${u<1}$ the dynamics that is generated by $\mathcal{H}$ 
is topologically the same as Rabi rotations:
the phase-space trajectories around $X$ are 
merely deformed. This means that there are two 
stable fixed-points, both located on the Equator (${\theta=\pi/2}$). 
The ground-state fixed-point is ${\varphi=0}$, 
and the upper-state fixed-point is ${\varphi=\pi}$. 
For ${u>1}$ the  ground-state fixed-point remains 
stable but the upper fixed-point bifurcates.
Instead phase-space can support condensation
in the North or in the South fixed-point.
See \Fig{f2} for illustration.  This is an example 
for the 2nd scenario that has been mentioned in the introduction.

{\bf Mott transition.--} The dimer constitutes a minimal model also 
for the demonstration of the Mott transition, 
which is the 3rd scenario that has been mentioned 
in the introduction. Namely, if we increase $u$ 
beyond $N^2$ the the area of the Rabi region 
in phase-space becomes smaller than Planck cell. 
This means that the ground-state is no longer 
a coherent state. Rather it becomes a Fock state 
with 50\%-50\% occupation of the two wells.
Semiclassically it is represented by a strip along 
the Equator, with uniform $\varphi$ distribution.

\begin{figure}
\centering

\hspace*{-5mm}
\includegraphics[width=1.1\hsize,clip]{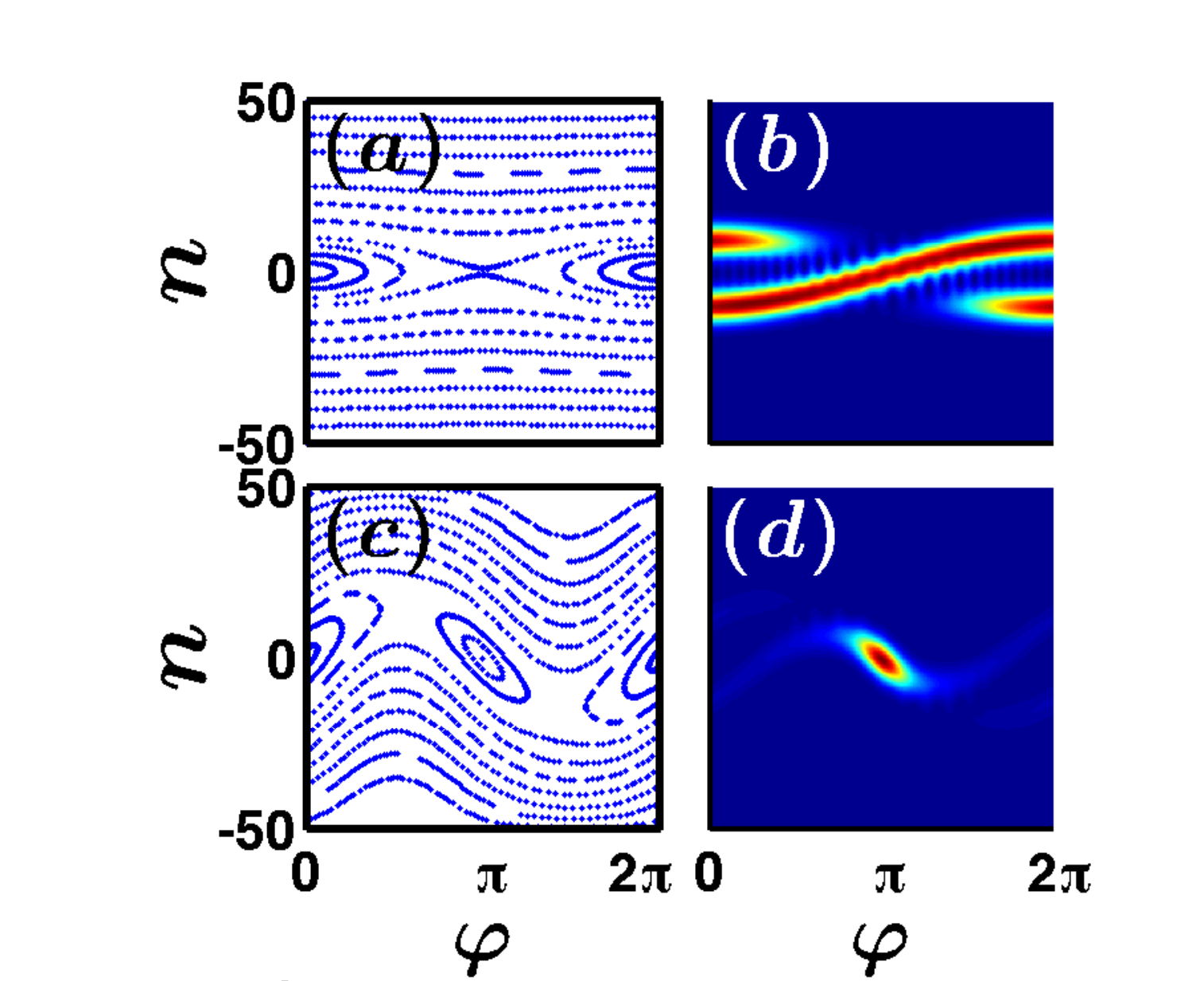}

\includegraphics[width=0.9\hsize,clip] {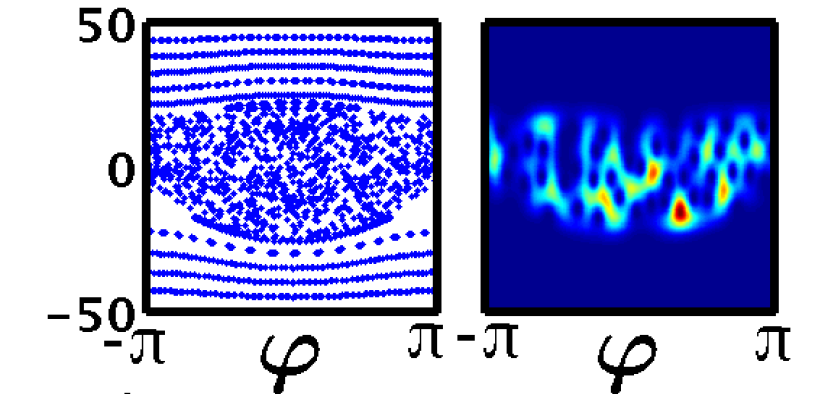}

\caption{
The dynamics of the dimer is illustrated in phase-space. 
For numerical details see \mycite{csk}.
The upper panels simulate the evolution 
in the absence of external driving:  
The left panel shows the classical phase-space 
portrait, and the right panel provides the Husimi 
representation of a quantum-mechanically 
time-evolved $\pi$ preparation 
(red means high probability).  
The 2nd row panels show what happens 
if one adds high frequency periodic driving 
that converts the hyperbolic fixed-point 
into a stable elliptic fixed-point.
The 3rd row panels show what happens if the 
driving frequency is comparable with 
the natural frequencies of the dynamics:
one observes a chaotic sea within which 
the quantum state ergodizes.  
} 

\label{f3}
\end{figure}

\begin{figure}
\centering

\includegraphics[width=\hsize]{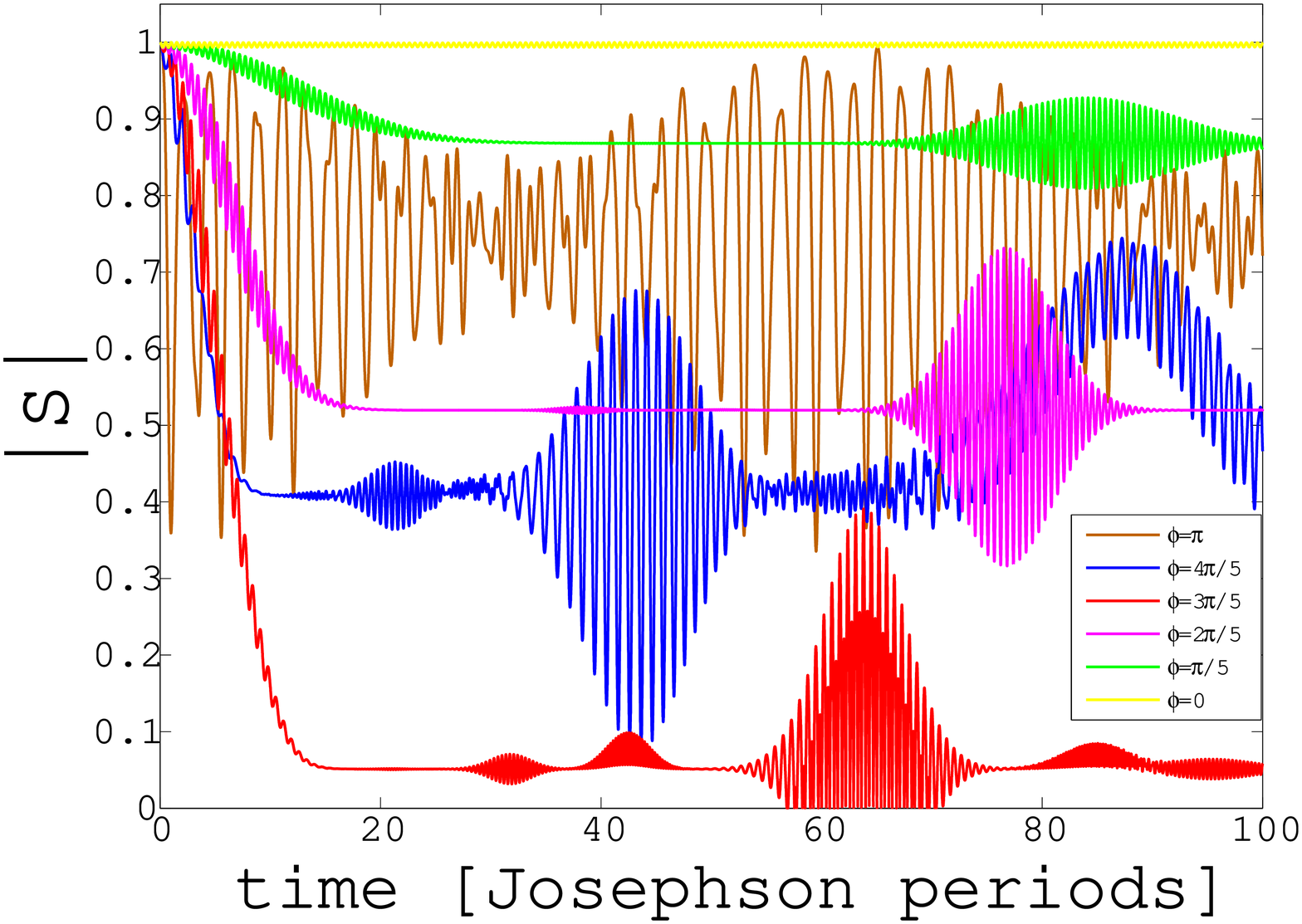}

\includegraphics[width=\hsize]{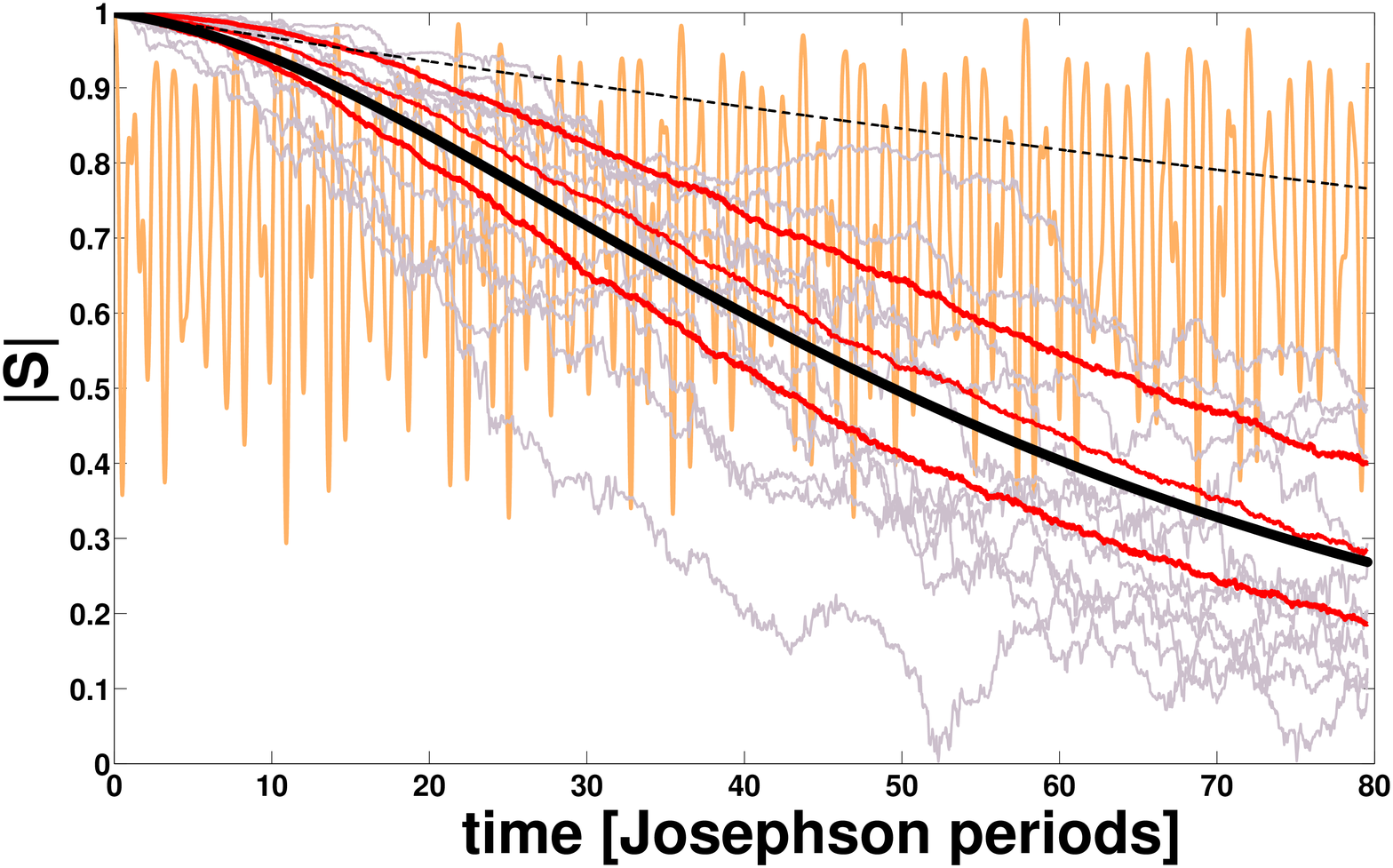}

\caption{
The length $|S|$ of the Bloch vector is a measure for 
one-body coherence. We follow the time evolution of 
a coherent state. In the upper panel each curve is for 
a different preparation (see text). In the lower panel
the evolution of a ${\varphi=\pi}$ preparation is simulated
in the presence of noise (thick black curve) and compared
with noiseless evolution (orange curve). 
For more details see \mycite{csm}. } 

\label{f4}
\end{figure}

\section{Quasi-stability of an unstable preparation}

Having figured out that for ${u>1}$ the ${\varphi=\pi}$
fixed-point is not stable, the question arises what happens 
if initially we condense all the bosons in the upper orbital. 
Semiclassically such preparation is represented 
by a Gaussian-like distribution at the ${\varphi=\pi}$ fixed-point, 
see \Fig{f2}b. It is useful here to make a connection 
with the Josephson Hamiltonian. Namely, in the vicinity 
of the Equator \Eq{e5} can be approximated by 
\beq
\mathcal{H}_{\text{Josephson}} \ \ =  \ \ U n^2 - \frac{NK}{2} \cos(\varphi)
\eeq   
where $\varphi=\varphi_1-\varphi_2$ is the conjugate phase. 
This is formally like the Hamiltonian of a mathematical pendulum.
The ${\varphi=\pi}$  preparation is like trying to position 
the pendulum in the upper unstable point. Our classical intuition 
tells us that such state should decay exponentially.
Using a phase-space picture, the wavepacket is expected to squeeze 
in one direction and stretch in the other (unstable) direction, along the separatrix.  
This is demonstrated in the upper panel of \Fig{f3} using 
the Husimi representation.

However, it turns out that the naive classical intuition 
with regard to the stability of the ${\varphi=\pi}$ 
preparation fails once longer time are considered. 
In the upper panel of \Fig{f4} we plot the time evolution of 
the length $S_B=|\bm{S}|$ of the Bloch vector 
for various coherent preparations
\beq
\bm{S} \ \ = \ \ \frac{2}{N} \left( \langle J_x \rangle, \langle J_y \rangle, \langle J_z \rangle  \right)
\eeq
Note that the reduced probability matrix 
is expressible in terms of $\bm{S}$, 
hence the purity measure that has been defined 
in the introduction is $S=(1+S_B^2)/2$.  
The initial length ${S_B{=}1}$ of the Bloch vector
reflects the coherence of the initial preparation.
If the initial state is located along the Equator 
at $\varphi=0$, it remains there (stable). 
If it starts elsewhere it typically decays. 
But if it starts at $\varphi=\pi$, 
the motion is dominated by recurrences: it becomes quasi-periodic, 
hence this preparation is quasi-stable.

\begin{figure*}
\centering

\mpg[58mm]{\includegraphics[height=5cm]{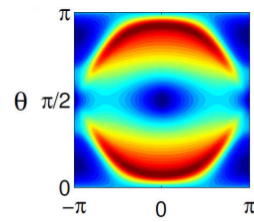} \\ \ \vspace*{1mm}  }
\includegraphics[height=0.3\hsize]{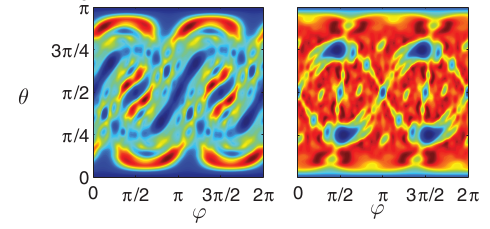}

\caption{
An image of $\text{PN}(\varphi,\theta)$.
The left panel, taken from \mycite{csf}, 
is for the integrable non-driven dimer 
with $N{=}100$ bosons, and $u{=}2.5$. 
The color code goes from PN$\sim1$ (deep blue) to PN$\sim30$ (red). 
The middle and right panels are for a kicked dimer 
that has a mixed or a fully-chaotic phase-space, respectively. 
For extra details see \mycite{ckt}. 
In the latter (chaotic) case one observes scarring (blue regions) 
in the vicinity of the classical hyperbolic points. 
} 

\label{f5}
\end{figure*}

In order to explain this quasi-stability we expand 
the initial coherent state in the basis of $\mathcal{H}$ 
eigenstates. Then we determine the {\em participation number} (PN) 
of the preparation. The PN tells how many eigenstates ``participate" 
in the superposition. If the superposition involved 
all the eigenstates we would get PN$\sim N$. 
For a coherent-state, which is like a minimal wavepacket,   
the naively expected result is PN$\sim1$ if it is located 
in the vicinity of a stable elliptic fixed-point, 
and PN$\sim \sqrt{N}$ otherwise. 
\Fig{f5} provided an image of PN for all possible coherent-preparation.
The color of a given point ${(\varphi,\theta)}$ 
reflects the PN of a ``minimal wavepacket" that is launched 
at that point. We see in the left panel that 
the PN of a $\varphi{=}\pi$ preparation is of order 
unity (color-coded in blue) contrary to the naive expectation.
An analytic calculation using a WKB approximation \mycite{csd,csf} 
provides the estimate PN$\sim \log(N)$. 
This explains the quasi-stability of the unstable fixed-point: 
the PN is typically small hence the motion is qusi-periodic.
A huge value of $N$ is required to get the irreversible decay 
that would be observed in the classical limit.

The low PN of a ``minimal wavepacket" that is launched 
at the vicinity of an hyperbolic point can be regarded 
as an extreme example for ``quantum scarring". The latter 
terms is reserved to the case where an hyperbolic point 
is immersed in a chaotic sea. 
\Fig{f5} provides an example for a PN calculation 
for a kicked dimer \mycite{csk}.
The middle panel is for a mixed phase-space system 
where the low PN regions simply reflect quasi-integrable motion.  
The right panel is for a strongly chaotic system 
where the blue regions indicates the presence 
of a classical hyperbolic point. Strangely enough in 
a classical simulation the hyperbolic point cannot be detected 
because it has zero measure. But quantum mechanics 
is generous enough to acknowledge its existence.    
We note that in this ``quantum scarring" example PN$\sim N$ 
for any preparation: the low PN is due to a prefactor
is the quantum scarring formula, and not due to a different 
functional dependence on~$N$.

\section{Stabilization - The Kapitza and the Zeno effects}

The ${\varphi=\pi}$ preparation is qusi-stable rather than stable. 
The question arises whether in an actual experiment it can be 
stabilized such that $S\sim1$ for a long duration of time.
The answer is positive. It can be stabilized by introducing 
hight-frequency or noisy driving. The Hamiltonian becomes 
\beq
\mathcal{H}_{\text{total}} \ \ = \ \ \mathcal{H} \ \ + \ \ f(t)W
\eeq 
The coupling is via some~$W$. In the present context 
we assume that the hopping amplitude $K$ is modulated, 
accordingly $W=J_x$.

By periodic driving we mean
\beq
f(t) \ \ = \ \ A\sin(\Omega t)
\eeq
One should be aware that if $\Omega$ is comparable 
with the natural frequency 
of the system, we merely get chaotic dynamics 
as demonstrated in the lower panels of \Fig{f3}.
This means that stability is completely lost.
But if we have high-frequency driving, its effect is 
averaged, and we get quasi-integrable motion with 
an effective Hamiltonian $\mathcal{H}+V^{\text{eff}}$, where
\beq
V^{\text{eff}} \ \ = \ \ -\frac{A^2}{4\Omega^2} \ [W,[W,\mathcal{H}]] 
\eeq
See \mycite{csk} for derivation. It turns out that the 
additional term converts the hyperbolic point 
into a stable elliptic point as demonstrated in the 
middle panels of \Fig{f3}. This is known as the Kapitza 
effect. We merely generalized here the standard analysis 
of the canonical mathematical pendulum.

By noisy driving we mean that $f(t)$ looks like ``white noise" 
with zero average and correlation function 
\beq
\overline{f(t)f(t')} \ \ = \ \ 2D \delta(t-t')
\eeq 
Using standard elimination technique one 
concludes that the dynamics is described by 
the following Fokker-Planck equation:
\beq
\frac{d\rho}{dt} \ \ = \ \ -i[\mathcal{H} ,\rho] - D [W,[W,\rho]]
\eeq
The analysis \mycite{csm} shows that the decay 
is described by the expression 
\be{13}
S_B \ \ = \ \ \exp\left\{-\frac{1}{N} \left[\exp\left(8D_wt\right)-1\right] \right\}
\eeq
where the radial diffusion coefficient is  
\beq
D_w \ \ = \ \ \frac{w_J^2}{8D}
\eeq 
The stronger the noise, the slower the radial diffusion.
Looking at \Fig{f4} we see that we have achieved $|S|\sim1$ 
for a long duration. In the remaining paragraphs of this 
section we shall provide a heuristic explanation for this effect.

There is related stabilization method that comes under 
the misleading title ``quantum Zeno effect".
The idea is to ``watch" the pendulum. Due to successive ``collapses" 
of the wavefunction the decay is slowed down. Using 
a standard Fermi-golden-rule analysis one deduces the expression
\be{15}
S_B \ \ = \ \ \exp\left\{-\frac{1}{N} 8D_w t \right\}
\eeq
This expression coincides with \Eq{e13} for very 
short times, and fails for longer times, 
as demonstrated in the lower panel of \Fig{f4}, 
where it is plotted as a thin black line.
What is misleading here, is the idea that the Zeno 
effect is a spooky quantum effect. In fact to ``watch"
a pendulum is formally the same as introducing noise.
The effect of the noise is to stabilize the pendulum, 
and this would happen also if Nature were classical...  

So what is the essence of the Zeno effect?  
The most transparent way to explain it is to use 
a phase-space picture. Let us regard the 
wavepacket as an ellipse with area ${\mathsf{A}=\pi r_a r_b}$. 
The effect of $\mathcal{H}$ is to squeeze it in one 
direction and stretch it in the other direction.
Note that the area $\mathsf{A}$ is not affected (Liouville's theorem).   
The effect of $f(t)J_x$ is to induce random rotation 
of its orientation. Thanks to the random rotations 
the stretching process is slowed down.
Schematically we can write the length of 
the the randomly rotating major axis as 
\beq
r(t) \ \ = \ \ \lambda_t \ ... \ \lambda_2 \ \lambda_1  \ r(0)
\eeq
where $\lambda$ is either smaller or larger than unity 
depending on the orientation of the ellipse. 
The net effect is diffusion of $\log(r)$, leading to \Eq{e13}.
Now we can also understand what is the reason for the 
failure of \Eq{e15}. The first-order treatment 
involves the substitution ${\lambda=1+\epsilon}$, 
and then the product is expanded. Hence $r(t)$ becomes 
a sum ${\epsilon_t + ...  + \epsilon_2 + \epsilon_1}$, 
rather than a product of random variables, and one 
deduces wrongly that $r(t)$ diffuses, leading to \Eq{e15}.

\begin{figure*}
\centering

\begin{minipage}{0.4\hsize}
\includegraphics[width=\hsize]{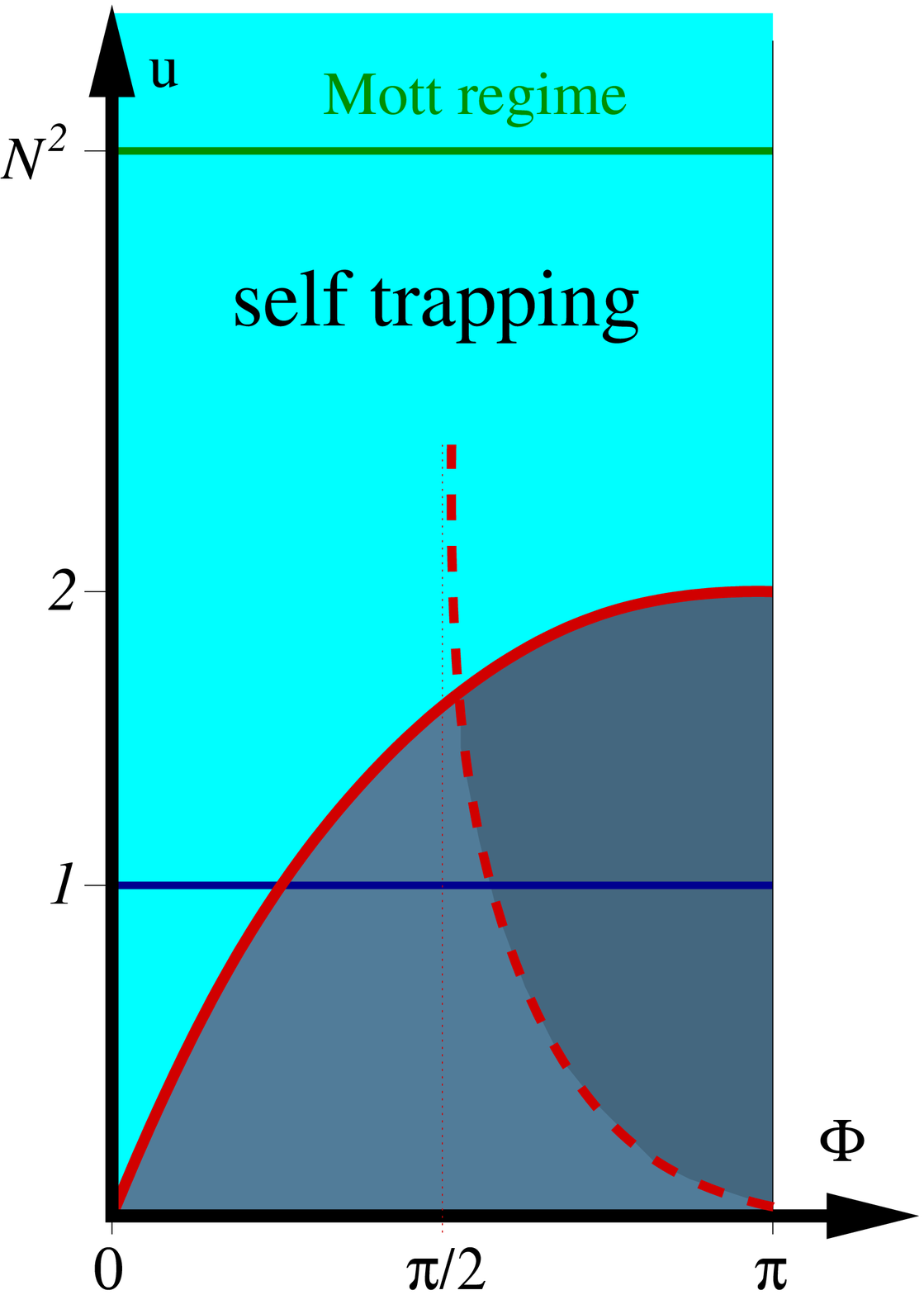}
\end{minipage}
\ \ \ \ \ 
\begin{minipage}{0.4\hsize}
\includegraphics[width=0.95\hsize]{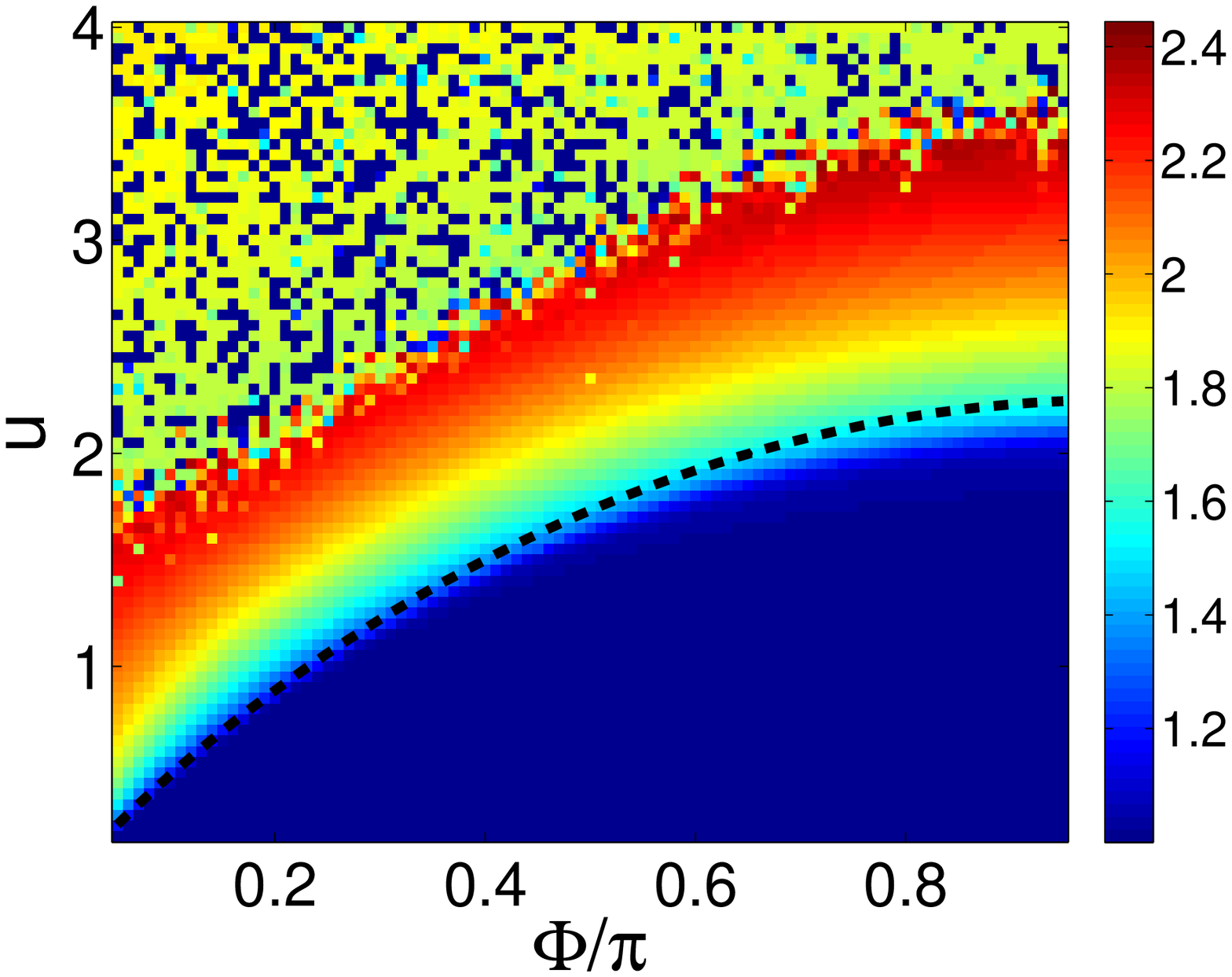} \\
\includegraphics[width=0.95\hsize]{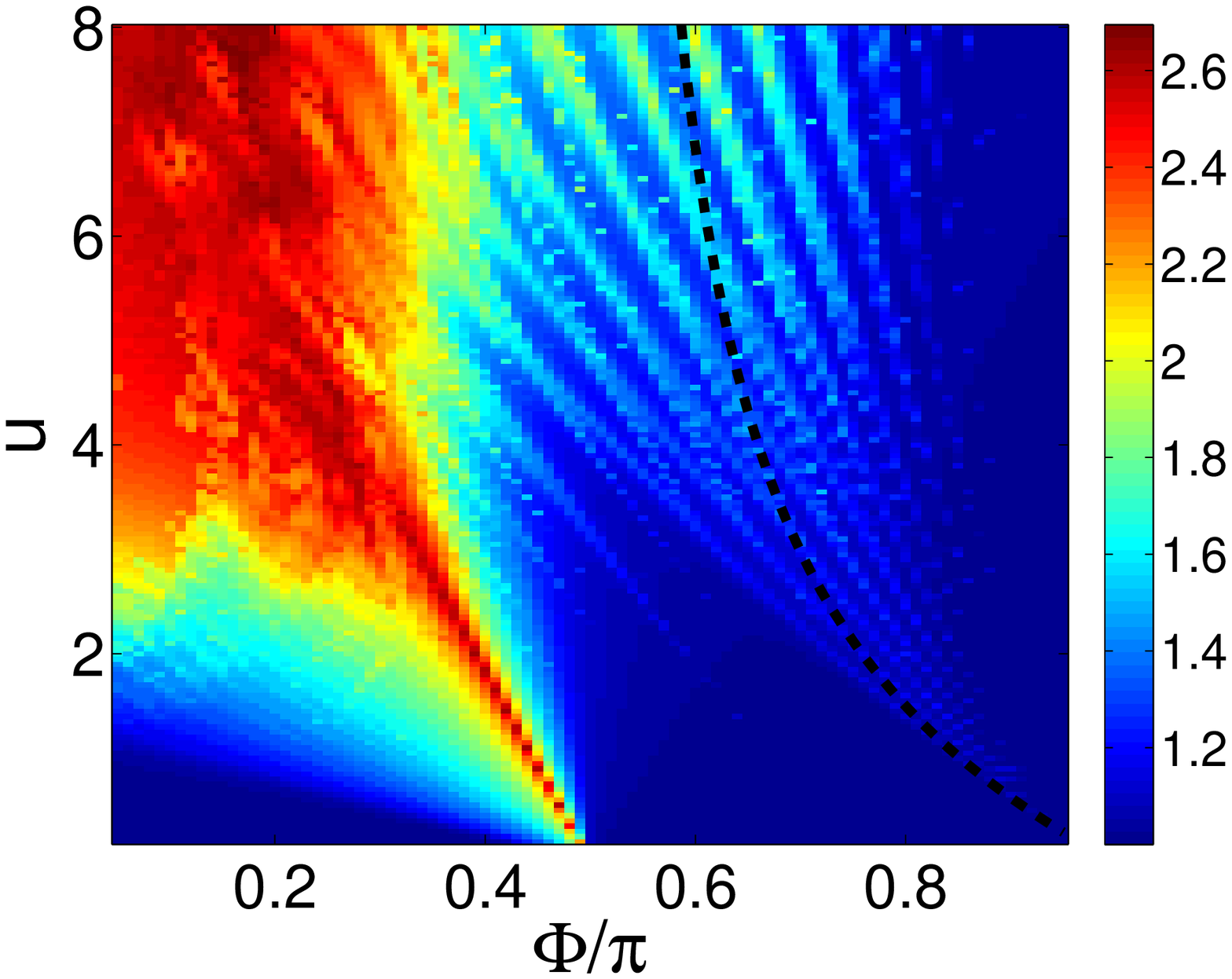}
\end{minipage}

\caption{
Regime diagram of a triangular BEC trimer with $N{=}42$ bososn, taken from \mycite{sfs}. 
The model parameters are $(\Phi,u)$.
{\em Left panel}: 
The thick red line is the semiclassical stability border 
of the upper vortex-state, where it bifurcates 
into 3  self-trapped states.
The thick dashed line is the Landau-criterion  
stability border of the intermediate vortex state:
the expected stability region is 
on the right of this curve. 
{\em Right panels}: 
The inverse purity $1/S$ is imaged 
for the upper most eigenstate (upper panel), 
and for the maximal current eigenstate (lower panel).
In the former case we observe an agreement 
with the classical stability analysis, 
while in the latter case we observe 
unexpected quasi-stability outside 
of the semiclassical stability region.
} 

\label{f6}
\end{figure*}

\section{The trimer - a minimal model for a superfluid circuit}

The BHH of the trimer is the same as \Eq{e3}, but for 
sake of generality we add in the kinetic term hopping phases 
that reflect the Coriolis field: 
\beq
-\frac{K}{2} \sum_{j=1}^{M} \left(
\eexp{i(\Phi/M)} a_{j{+}1}^{\dag} a_{j} + \eexp{-i(\Phi/M)} a_{j}^{\dag} a_{j{+}1} 
\right)
\eeq
Note that this is formally like having an Aharonov-Bohm 
magnetic flux through the ring. 
The classical energy landscape of the BHH 
always has a lowest fixed-point that might support a vortex-state, 
and an upper fixed-point that might support either a vortex-state 
or (due to bifurcation) a set of self-trapped states. 
Note that the upper-state can be regarded as a ground-state of 
the ${U\mapsto -U}$ Hamiltonian. 
The $(\Phi,u)$ regime diagram of this model is displayed 
in the left panel of \Fig{f6}.
As in the case of the dimer we have here two familiar 
scenarios: With regard to the {\em ground-state}, 
if $u$ becomes larger than $N^2$ it undergoes a Mott-transition  
and looses its purity (green line in \Fig{f6}). 
With regard to the {\em upper-state}, if $u$ crosses 
the solid red line, it bifurcates,  
and replaced by an quasi-degenerate set of 3~self-trapped states.

The inverse purity $1/S$ of the upper state is imaged 
as a function of $(\Phi,u)$ in the right upper panel of \Fig{f6}. 
The dashed line is the classical stability border of the 
vortex-state beyond which we have self-trapping. 
Clearly the numerical results agree with the classical prediction. 

The question arises whether it feasible 
to find a (meta)stable vortex-state, 
that is immersed in the ``continuum".
The ``continuum" is formed of states that are supported 
by the chaotic sea.  There are two possibilities here: 
The traditional possibility is to have a state that is supported 
by a stable fixed-point in an intermediate energy; 
The exotic possibility is to have quasi-stability 
in the vicinity of an unstable fixed-point. 
Looking at the right lower panel of \Fig{f6} we find 
that superfluidity survives beyond the classical (Landau) border of stability.
In particular we observe that superfluidity is feasible 
for a non-rotating device ($\Phi{=}0$), contrary to 
the traditional expectation that is based on the classical stability analysis.  
The full explanation of the superfluidity regime-diagram requires 
a thorough ``quantum chaos" analysis 
of the underlying mixed phase-space structure \mycite{sfc}. \\

\section{Concluding remarks}

The dimer is a minimal model for demonstrating the classical 
instability that leads to self-trapping, and the quantum Mott 
transition of the ground-state. It also provides an illuminating 
example for quasi-stability at the vicinity of an unstable 
hyperbolic point. Classical stabilization is feasible by 
introducing high-frequency periodic driving (Kapitza effect) 
or noise (Zeno effect). 

From topological point of view a triangular trimer
is the minimal model for a superfluid circuit. Due to the 
extra degree of freedom this model is no longer integrable, 
unlike the dimer. The question arises whether such 
circuit can support a metastable vortex-state. 
This is what one call ``superfluidity". 
We find that a quasi-stable superfluid motion manifests itself 
beyond the regime that is implied by the traditional stability analysis. 
The full explanation of the superfluidity regime-diagram 
of a low-dimensional circuit requires a thorough ``quantum chaos" analysis 
of the underlying mixed phase-space structure.

Low dimensional superfluid circuits 
are of current experiment interest \mycite{trmexp}. 
Periodically driven BEC circuits and the 
stabilization of nonequilibrium condensates 
is of special interest too \mycite{trmdrv,trmstb}.


{\bf Acknowledgements.-- }
The current manuscript is based on studies \mycite{csd,csf,ckt,csk,csm,sfs} 
that have been conducted in Ben-Gurion university 
in collaboration with my colleague Ami Vardi.
The main work has been done by Geva Arwas, 
Christine Khripkov, Maya Chuchem, and Erez Boukobza. 
The present contribution has been supported by 
the Israel Science Foundation (grant No.29/11).


\clearpage

\begin{thebibliography}{10}


\bibitem{BHH1}
O. Morsch, M. Oberthaler, 
Rev. Mod. Phys. 78, 179 (2006); 
%
I. Bloch, J. Dalibard, and W. Zwerger, 
Rev. Mod. Phys. 80, 885 (2008).


\bibitem{trimer5}
R. Franzosi, V. Penna,
Phys. Rev. A 65, 013601 (2002);
%
Phys. Rev. E {\bf 67}, 046227 (2003).

\bibitem{trimer11} 
M. Hiller, T. Kottos, and T. Geisel, 
Phys. Rev. A {\bf 79}, 023621 (2009).



\bibitem{Udea} 
R. Kanamoto, H. Saito, M. Ueda,
Phys. Rev. A 68, 043619 (2003),

\bibitem{Altman}
A. Polkovnikov, E. Altman, E. Demler, B. Halperin, M.D. Lukin, 
Phys. Rev. A 71, 063613 (2005)


\bibitem{Carr1}
R. Kanamoto, L.D. Carr, M. Ueda,
Phys. Rev. Lett. 100, 060401 (2008)

\bibitem{Brand2} 
O. Fialko, M.-C. Delattre, J. Brand, A.R. Kolovsky,
Phys. Rev. Lett. 108, 250402 (2012)

\bibitem{Ghosh}
P. Ghosh, F. Sols,
Phys. Rev. A 77, 033609 (2008) 



\bibitem{scars2}
L. Kaplan, E.J. Heller, 
Phys. Rev. E {\bf 59}, 6609 (1999).



\bibitem{csd} 
M. Chuchem, K. Smith-Mannschott, M. Hiller, T. Kottos, A. Vardi, D. Cohen,
Phys. Rev. A {82}, 053617 (2010).

\bibitem{csf} 
C. Khripkov, D. Cohen, A. Vardi,
J. Phys. A {46}, 165304 (2013).  

\bibitem{ckt}
C. Khripkov, D. Cohen, A. Vardi,
Phys. Rev. E {87}, 012910 (2013).

\bibitem{csk} 
E. Boukobza, M.G. Moore, D. Cohen, A. Vardi,
Phys. Rev. Lett. {104}, 240402 (2010).

\bibitem{csm} 
C. Khripkov, A. Vardi, D. Cohen,
Phys. Rev. A {85}, 053632 (2012); 
%
Eur. Phys. J. Special Topics {217}, 215 (2013). 

\bibitem{sfs} 
G. Arwas, A. Vardi, D. Cohen,
Phys. Rev. A (2014).

\bibitem{sfc}
G. Arwas, A. Vardi, D. Cohen (in preparation).

\bibitem{more} For a more comprehensive list of references 
see \mycite{csd,csf,ckt,csk,csm,sfs}, 
in particular \mycite{csd} (dimer) and \mycite{sfs} (trimer). 

\bibitem{trmexp}
L. Amico, D. Aghamalyan, F. Auksztol,H. Crepaz, R. Dumke, L.C. Kwek,  
Sci. Rep. 4, 4298 (2014).

\bibitem{trmdrv}
M. Heimsoth, C.E. Creffield, L.D. Carr, F. Sols, 
New Journal of Physics 14 075023 (2012)

\bibitem{trmstb}
B. Gertjerenken, M. Holthau, 
arXiv:1410.8008

\end{thebibliography}
\end{document}